# Variation of $TiO_2/SiO_2$ mixed layers induced by different $Xe^+$ ion energies

*Tran Van Phuc[1,2,3*], Miroslaw Kulik[3,4], Afag Madadzada[3,5], Dorota. Kołodyńska[6], Le Hong Khiem[1,2], Phan Luong Tuan[3,7], Nguyen Ngoc Anh[8]*

[1]*Graduate University of Science and Technology, Vietnam Academy of Science and Technology, 18 Hoang Quoc Viet, Cau Giay, Ha Noi, Viet Nam,*

[2]*Institute of Physics, 10 Dao Tan, Ba Dinh, Ha Noi, Viet Nam,*

[3]*Frank Laboratory of Neutron Physics, Joint Institute for Nuclear Research, 141980 Dubna, Russia,*

[4]*Institute of Physics, Maria Curie-Skłodowska University, M. Curie-Skłodowska Sq. 1, 20-031 Lublin, Poland,*

[5]*Department of Neutron Physics, National Nuclear Research Centre JSC, Baku-Shamahi hw 20km, AZ 0100 Baku, Azerbaijan.*

[6]*Institute of Chemical Sciences, Faculty of Chemistry, Department of Inorganic Chemistry, Maria Curie-Skłodowska University, M. Curie-Skłodowska Sq.2, 20-031, Lublin, Poland,*

[7]*Hanoi Irradiation Center, Vietnam Atomic Energy Institute, 59 Ly Thuong Kiet, Hoan Kiem, Hanoi, Vietnam*

[8]*Institute of Research and Development, Duy Tan University, Danang 550000, Viet Nam.*

*\* Corresponding author*

*E-mail: tvphuc@iop.vast.ac.vn*

## Abstract

The broadening and optical parameters of $TiO_2/SiO_2$ transition layers depending on the ion energy have been investigated using the Rutherford Backscattering Spectrometry (RBS) and Ellipstrometry Spectroscopy (ES) methods. The $TiO_2/SiO_2$ samples were irradiated by $Xe^+$ ions with energies of 100, 150, 200 and 250 keV. The depth profiles of the elements determined by the RBS spectra show the structure and thickness of the $TiO_2/SiO_2$ transition layers before and after implantation. We have found that the thickness of the transition region between the $TiO_2$ and $SiO_2$ layers increases with the increasing incident ion energy. This phenomenon indicates an increasing amount of atomic mixing at the $TiO_2/SiO_2$ interface. In addition, the variation of transition layers could be explained by defect depth profiles and ions energy transferred in the mixed layers by means of SRIM calculations. The thickness obtained from the RBS is in good agreement with that measured using the ES method. Based on these obtained results, we have also investigated the optical constants of implanted and non-implanted $TiO_2/SiO_2$ structures. The wave forms measured with varying incident angles suggest that the measurements were made close to near the main principle point. The yields of $\Psi(\lambda)$ and $\Delta(\lambda)$ bands vary at different incident angles, is associated with interference processes of the light reflected from the structures examined. The refractive index and the extinction coefficient were found to increase after implantation taking place up to 200-keV Xe and then decrease at 250 keV.

**1. Introduction**

It has been well known that structures and properties of many materials are able to be modified in a controlled way by applying ion implantation [1]. The process of ion implantation gives two main results that is transferring atoms and their energy change into the material. The *ion beam mixing* (IBM) is an important characteristic of this process. The IBM phenomenon is caused by the interactions of ions and target atoms at the interface of two adjacent material layers [2].

When the transferred energy exceeds the displacement energy ($E_d$) of a target atom, this atom is displaced from its lattice site to a new location in the neighboring layer. Then, a secondary interaction can take place, lead to a collision cascade which results in formation of a mixed area between those materials. Applying this mechanism, modification of system properties of system properties is possible by ion beam mixing in the ways that are difficult or impossible to achieve by conventional methods. IBM became a powerful tool commonly used the formation of stable, metastable, amorphous and crystalline phases in the bilayer and multilayer [3,4]. Consequently, it is a great importance in material science and technology to seek for deeper understanding of the interaction processes through the interfaces and the formation of properties induced by IBM. In fact, potential applications and fundamental mechanism of the IBM phenomenon in various structures systems, including metal-metal [5,6], metal-silicon [7] and metal- insulator [8] systems, have been studied.

Several advantages such as chemical and thermal stability, low cost, electronic properties and long durability make $TiO_2$ the most studied material among different light harvesting materials used as photo anodes in the photoelectrochemical (PEC) cells [9,10]. $TiO_2$ also possesses a wide range of applications in the decomposition of organic pollutants, self-cleaning coatings, photovoltaics, biomedical devices, electrochromic display devices and Li-ion batteries [11, 12]. However, there are also a few undesired characteristics such as the large band-gap and fast recombination of electrons and holes, which lessen the application of $TiO_2$ [13]. Coupling with $SiO_2$ is shown as a method that can enhance the $TiO_2$ activity due to the increase in the adsorption amount of substrates and the improvement of dispensability in water [14]. On the other hand, Fernández et al. reported the electronic support interactions induced by the interfacial Si–O–Ti bonds between the $TiO_2$ thin film and the $SiO_2$ substrates [15]. However, the study on the $TiO_2$–$SiO_2$ system focusing on the interfacial interactions is limited and the essential origin of the coupling effect is not fully understood. This paper presents the results of investigations of Xe implantation with different energies into the $TiO_2/SiO_2$ bilayers covering on the Si substrates. The process of forming mixed layers of the oxide-oxide systems proceeds by means of RBS [16] and SE [17] methods. The effects of Xe ions implantation with different energies on broadening of the mixed layers, their optical properties and chemical compositions were determined.

## 2. Experimental

In this study the $TiO_2/SiO_2$ bilayers deposited on the Si substrates were investigated. The specimens were irradiated with the $Xe^+$ ions at different energies of 100, 150, 200, 250 keV. The radiation was carried out at room temperature with the fixed value of fluence at $3 \times 10^{16}$ [atoms/cm$^2$] using the ion implanter UNIMAS at the disposal of Maria Curie-Skłodowska University [18]. The elemental depth profiles of the samples were determined using the RBS method. The 1.5 MeV $He^+$ ion beam was used with a 0.5 mm beam diameter. It was directed towards the samples at the incident angle $60^0$ towards the normal of the sample surface. A semiconductor detector with energy resolution about 12.76 keV was positioned at a scattering angle of $170^0$ away from the beam incident direction in order to collect the RBS spectra. The elemental composition and the depth distributions were obtained using the SIMNRA code [19]. The depth-dependent damage and defect concentration profiles were calculated for understanding and explanation of the obtained effects using the Stopping and Range of Ions in Matter (SRIM)-2008 [20]. The ions distribution, displaced atoms concentration and energy loss of the projectile in the transition layers were simulated using the detailed calculations with the full damage cascades option. The Xe ions were used due to the fact that they would not form any chemical binding with the target atoms during the interactions. Thus only physical structure of the samples was modified. For this kind of ion the energy was chosen so that they could interact with the atoms at both above and below the $TiO_2/SiO_2$ interface. The projected ranges of the Xe ions in the samples with different irradiation energies of 100, 150, 200 and 200 keV are shown in Fig.1. In addition, for the experimental analysis of optical properties of $TiO_2/SiO_2$ system, the ES measurements were performed using the rotating-analyzer ellipsometry (RAE) [14]. The ellipse of the angles $\Psi(\lambda)$ and $\Delta(\lambda)$ was measured with the light wavelength from 250 nm to 1100 nm, with the step of 1 nm at the angles of $70.0^0$, $72.0^0$, $74.0^0$, $76.0^0$ $78.0^0$ and $80.0^0$. From the spectra of $\Psi(\lambda)$, $\Delta(\lambda)$ recorded after the experiment and calculated using the MAIE method [21], optical constants, thickness of layers including the transition area and the content of $TiO_2$ and $SiO_2$ compounds in the layers before and after ion implantation were determined. The ES experiments were conducted at the Institute of Electron Technology, Lotników, Warsaw, Poland.

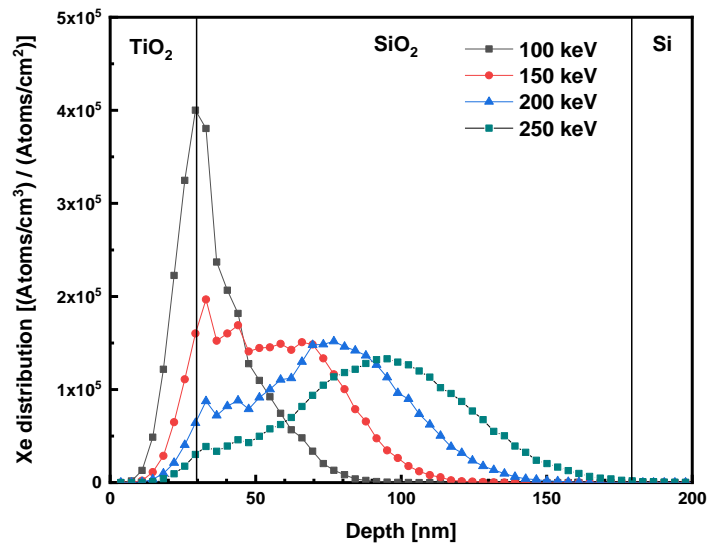

Fig.1. The projected ranges of the Xe ions in the samples with different energies of 100, 150, 200 and 200 keV (SRIM calculation).

## 3. Results and discussion

*3.1. Description of mixing process at the $TiO_2/SiO_2$ interface by the RBS method*

Fig.2 shows the typical RBS spectra collected from the samples implanted with $Xe^+$ ions at different energies. The spectra reveal the energy of backscattered ions versus their yield. In which a RBS spectrum shows overlapping of the Gaussian peaks that indicates the sum of near-surface layers thickness. During the ions collision with the near-surface atoms they could be backscattered with the same energy and collected around the same channel. The signal near such a channel associated with the kinematic borders (high energy edges) and indicated by the vertical arrows. Kinematic borders are spread only by the sputtering effect or by the resolution of the RBS detector. The inclined arrows in the figure indicate the borders that refer to the ions backscattered on the atoms in the subsurface layers. These borders could be shifted by changing the incident angle of the ion beam. Moreover, the back borders (low energy edges) could be shifted due to the contribution of the energy straggling, layers thickness and detector resolution. The width of the peaks or bands confined by the front and back borders is associated with the film thickness. There are several notable features from the spectra that deserve further attention. The band at the energy between 370 and 550 keV indicates the alpha particles backscattered from O in both $TiO_2$ and $SiO_2$ layers. The borders at the energy of 680, 820 and 1080 keV indicate silicon atoms in the $SiO_2$ layers, Si in the substrates and the Ti atoms in the $TiO_2$ layers, respectively. The presence of Xe ions in the samples after irradiation corresponds to the peaks that are the Gaussian distribution at the energy around 1270 keV. Existence of Xe atoms leads to a decrease of O and Si concentrations. This effect is associated with a significant decrease of the yield of related to O and Si near the energy of 480 and 750 keV, respectively. The spectra reveal a shifting position of the borders with the increasing ion energy. Such a shift can be partly attributed to the variation of Xe distribution as well as the difference in layer thickness of the implanted structures compared with that of the virgin samples.

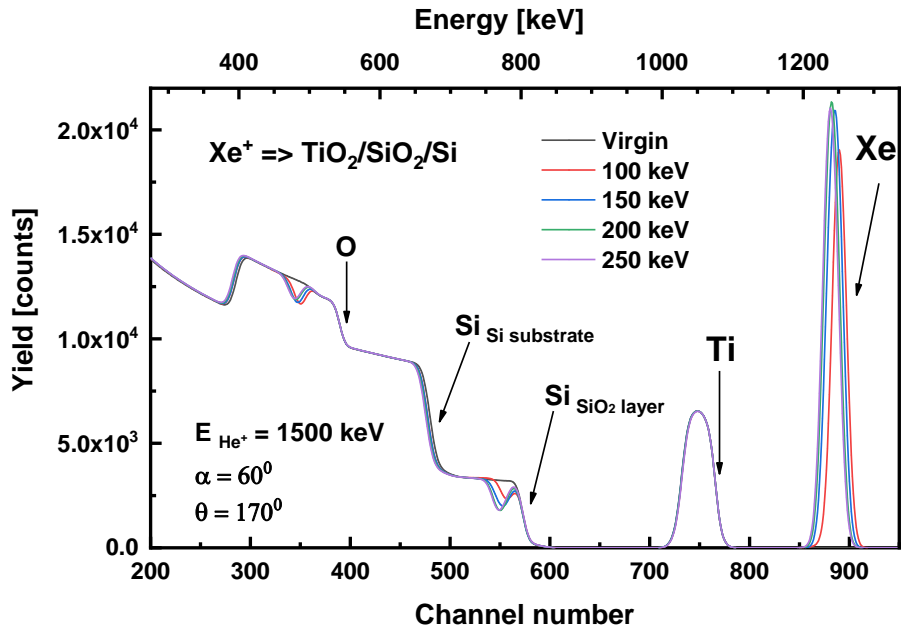

Fig.2. The RBS spectra collected from the samples before and after implantation with $Xe^+$ ions at different energies

The typical depth profiles for the elements in the samples implanted with 250-keV $Xe^+$ are shown graphically in Fig.3. The structure of $TiO_2/SiO_2/Si$ with the transition layer between them was determined based on the atomic distribution. $TiO_2$ and $SiO_2$ compounds could be distinguished due to the atomic concentration of Ti:O and Si:O at ratio of 1:2. The model layers making up the samples which were calculated using the SIMNRA code were assumed to be homogeneous. Thus the thickness of $TiO_2$ and $SiO_2$ layers was determined to be about 30 nm and 138 nm, respectively. For 250-keV $Xe^+$ ions depth profiles were obtained with an accessible depth at around 609 [nm] with accuracy of 0.1 [%] for Ti. After the ion irradiation, reduction in the concentration of Si and O can be observed at the depth from 60 nm to 80 nm. The silicon atoms amount occupied by $Xe^+$ ions increase while those of oxygen atoms decrease with the increasing energy of implanted ions. It was found out that mixing of $TiO_2/SiO_2$ interface indicated by the intersection of Ti and Si concentration lines in the range of 29 and 41 nm. In order to determine a mixing degree of transition layers towards the surface and substrate, changing the thickness of $TiO_2$ layers and transition layers as a result of ion beam irradiation was investigated in detail which is discussed below.

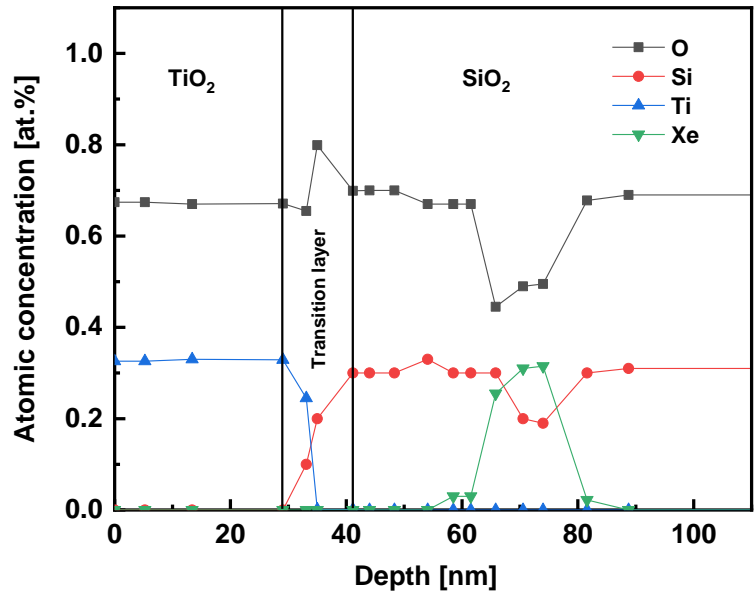

Fig.3. The depth profiles for the elements in the samples implanted with 250-keV $Xe^+$ ions.

The mixing process of the atoms across the $TiO_2/SiO_2$ interface is indicated by shifting the high energy edges of Si in the $SiO_2$ layers (Fig.4a) and the low energy edges of Ti in the $TiO_2$ layers (Fig.4b). When the ion energy increases, the low energy edges of Ti shift to the higher energy range. This regards decreasing Ti concentration in the tail of $TiO_2$ layers that make narrowing of initial $TiO_2$ and lead to broadening of the mixed layers towards the surface. The atoms in the mixed area could be also displaced to the $SiO_2$ layers resulting in broadening of the mixed layers towards the substrate which is indicated by shifting of the high energy borders of Si. Assuming that the mixing amount could be quantified by the edges shifting degree, influence of Xe energy on the mixing could be evaluated by determining the variation width of the Gaussian Ti peaks for the samples before and after implantation. The edges of Ti atoms were chosen due to great separation of Ti peaks from the Xe, O and Si signals in the RBS spectra. Changing thickness of $SiO_2$ layers was not investigated due to the variation at both $TiO_2/SiO_2$ and $SiO_2/Si$ interfaces of these layers.

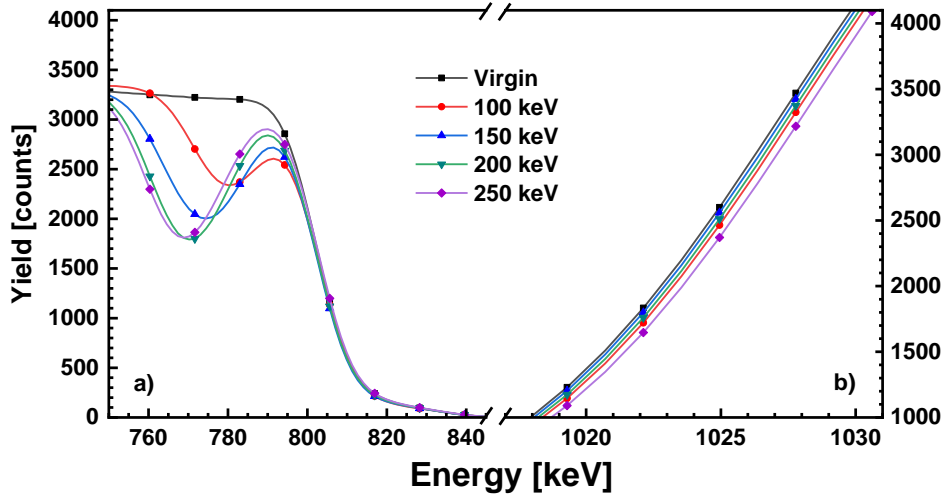

Fig.4. The fragments of RBS spectra for the high energy edge of Si in the $SiO_2$ layers (a) and low energy edges of Ti in the $TiO_2$ layers (b)

In the RBS spectra the signals are the sum contributions of the Gaussian distribution peaks collected from different thin layers. The peak shapes and positions are changed due to the ion energy loss, energy straggling and instrumental resolution. The low energy edges of the Ti spectra were determined by fitting the backscattering simulation for an error-function-like concentration profile [22]

$$c(x) = \{1 - erf[(x-d)/\sqrt{2}\sigma]\} \times c_0/2 \qquad (1)$$

where: $x$ is the location of the interface. In this equation use the spreading standard deviation $\sigma$ is used as a fitting variance. If we consider ions mixing as a diffusion process, D is the inter-diffusion coefficient at the interface and t is the effective time of the mixing process, $\sigma$ corresponds to the $Dt$ via $\sigma^2 = 2Dt$, $\sigma^2$ denotes the variance of the profile. The contribution of the ion energy to broadening of the mixed layers was deduced by subtracting $\sigma^2$ of the un-irradiated spectrum $\sigma_0{}^2$ from the variance $\sigma_{irr}^2$ after irradiation $\Delta(\sigma^2) = \sigma_{irr}^2 - \sigma_0{}^2$. Moreover, the standard deviation is related to the FWHM of a Gaussian distribution by FWHM = 2.355σ [16], thus $\Delta(FWHM^2) = (\text{FWHM})_{irr}^2 - (\text{FWHM})_0^2 = 2.355^2 \, (\sigma_{irr}^2 - \sigma_0{}^2)$. These analysis shows that $\Delta(FWHM^2)$ gives information about the effective diffusion coefficient while $\Delta_{FWHM}$ refers to changing the Ti concentration in the tail of the TiO$_2$ layers after implantation.

With increasing of the implanted ion energy, decreasing FWHM of the peak Ti was found. This appropriated to reducing Ti concentration in the tail of TiO$_2$ layers that make the broadening of the transition layer towards the surface. Fig.5a shows the variation of $\Delta_{FWHM}$ Ti peaks of the RBS spectra collected from the samples before and after implantation with Xe ions as a function of the ion energy. The curve reveals clearly the largest decrease of the $\Delta_{FWHM}$ for the case of 250-keV Xe implantation, followed by 100, 200 and 150 keV. For 100 keV Xe, FWHM decreases (0.8%) more stronger than for 150 kev Xe (0.3%) due to a larger number of ions concentrated near the interface producing greater defects. While the 150-keV Xe ion penetrates more deeply, thickness of TiO$_2$ layer is less affecting by collisions and a smaller, less decrease of $\Delta_{FWHM}$ could be observed. The peak width decreases more for 200 keV Xe implantation (0.7%), followed by the greatest drop for 250 keV (1.6%) due to the strong effects induced by the higher transfer ion energy. During the interactions with ions, the atoms in the tail of TiO$_2$ layers are displaced, as a result the TiO$_2$ layers are narrowed and leads directly to reduction of defects amount. The number of atoms sputtering from TiO$_2$ was assumed to be insignificant. Thus the reducing degree of the $\Delta_{FWHM}$ could be associated with amount of the displacement atoms and vacancies in the TiO$_2$ layers (Fig.5b). A high value of defects could be seen for the case 100 Xe due to a short projected range of ions which makes them interact near the tail of TiO$_2$ layers. Then the number of defects decreases linearly with the increasing of ion energy that could be explained by infiltration of ions towards the SiO$_2$ layer. Since the ions produce the defects above and below the TiO$_2$/SiO$_2$ interface in the energy range of [100-250] keV, the thickness of the mixed layers was changed by defects found at the front and back borders of the transition layers. Thus the variation of the TiO$_2$ layer thickness $\Delta_{FWHM}$ lacks to qualify the mixing amount, the changing thickness of the mixed layers compared with the initial transition area must be taken into account. The determination of the changes in both the TiO$_2$ and the transition layer thickness could give comparison broadening of the transition layer at the upper and lower borders of the TiO$_2$/SiO$_2$ mixed area.

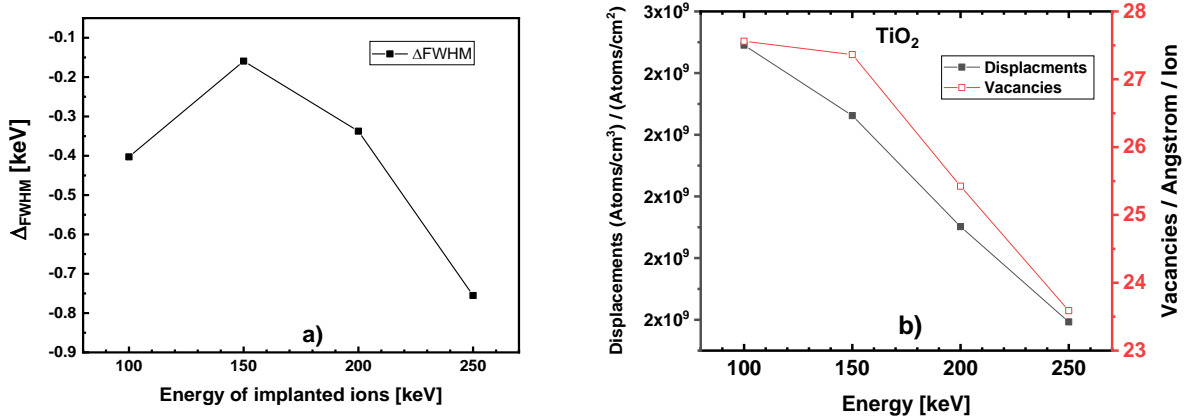

Fig.5. The variation of $\Delta_{FWHM}$ of Ti peaks from the RBS spectra (a) and damages amount in the TiO$_2$ layers calculated by SRIM (b) as the function of Xe ion energy.

The variation in the transition layer thicknesses of TiO$_2$/SiO$_2$ structures before and after the ion irradiation were determined using the elemental depth profiles. The relative thickness which refers to the degree of changing width of mixed layers was defined from the equation $r_t = (t_{im} - t_{vir})/t_{vir}$, where $r_t$ is the relative thickness of the layers, $t_{vir}$ and $t_{im}$ are the thickness of the layers in the implanted and non-implanted materials, respectively. The role of the Xe ion energy in the mixing process was studied by evaluating $r_t$ at different energies from 100 to 250 keV. The $r_t$ values show to be increase with increasing of energy, this behavior corresponds with variation of the energy transfer to the recoil atoms in the transition layers as shown in the Fig.6a.

Growing of the energy to recoil points out that the atoms near interface were transferred by higher energy, they were displaced, travelled with the longer distance from the interface then the thicker transition layers were caused. In the light of Sigmund's conclusion that the noticeable increase in mixing rates occurs at a fixed depth with increasing energy [23], this effect could be explained. The results indicated a strong broadening of mixed area for 100 keV Xe compare with that of virgin sample. The mixing proceeds as the energy increase up to 200 keV, then slowly increase for the highest energy of the Xe ion. After implantation, the collision produces displaced atoms at mixed area, their concentration depends on the ion beam parameter as well as the target state. The mixing degree corresponds to the variation of $r_t$ thus could be referred to the number of atoms in the implanted transition layers. By means of RBS and SRIM calculation, surface density of the atoms and total number of displacements in the transition layers was determined. Fig.6b shows variation of these parameters as a function of the ion energy, a similar increasing trend was also observed as a consequence of increasing thickness of the transition layers.

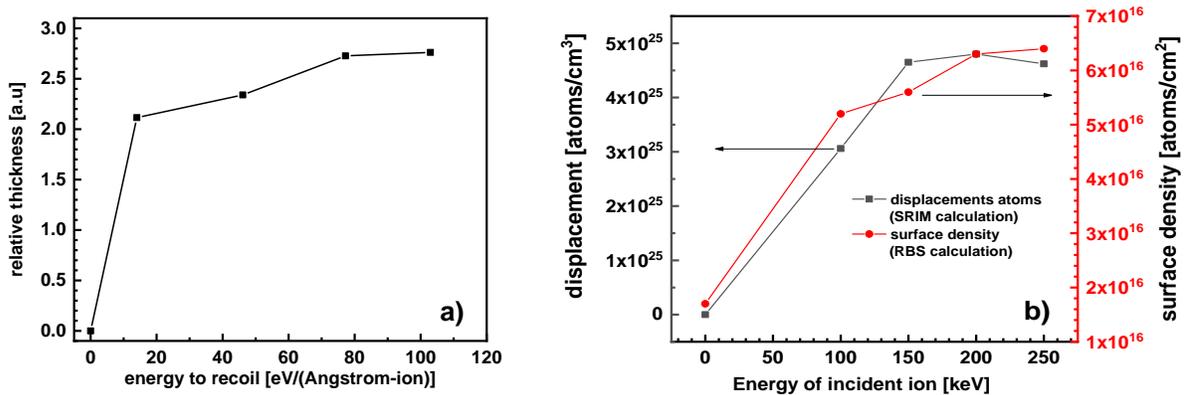

Fig.6. The variation of relative thickness versus energy to recoil (a) and increasing the surface density and displacements versus of ion energy (b).

This effect could be interpreted by the increasing in the distance of the ions from the interface of $TiO_2/SiO_2$ when the ion energy grown. For 100-keV $Xe^+$, the projected range (the peak ion concentration) is determined to be around 28.7 nm (Table 1) near the interface, where the ions interact, slow down and stops by transferring all of the energy to the atoms. For higher energy, the projected ranges increase which lead to reducing the number of interacting ions in the mixed layers, makes decreasing the intensity of atomic displacement in this region. In the energy range between 100 and 250 keV, electronic stopping which is caused by the interaction between the incoming ion and the electrons in the target is dominant.

Table 1. The ion projected ranges and the transport features of the ions in the $TiO_2/SiO_2$ transition layers.

| Incident ion energy [keV] | Projected range [nm] | Ions density in mixed layers [$10^{21}$×atoms/cm$^3$] | Energy to recoil [eV/(Å*ion)] | Energy loss [eV/(Å*ion)] | |
|---|---|---|---|---|---|
| | | | | Nuclear | Electronic |
| 0 | - | 0 | 0 | 0 | 0 |
| 100 | 28.7 | 30.6 | 14.00 | 3.50 | 21.68 |
| 150 | 57.4 | 15.3 | 46.09 | 4.53 | 53.30 |
| 200 | 71.7 | 6.7 | 77.22 | 4.34 | 80.63 |
| 250 | 90.8 | 3.1 | 102.94 | 3.99 | 102.53 |

It was noticed that increasing transition layers is associated with decreasing $\Delta_{FWHM}$ of Ti peaks. However, the mixing process occurs in two directions that toward the substrate and the samples surface. Based on the depth profiles of elements determined by RBS, the obtained average ratio of Si/Ti atoms in the transition layers is 2.4. Moreover, the thickness of the mixed layers after irradiation increases 3.1, 3.3, 3.5 and 3.8 times higher than that of the initial transition layers for 100, 150, 200, 250 keV, respectively. Compared with 0.8%, 0.3%, 0.7% and 1.6% decreasing $\Delta_{FWHM}$, it could be pointed out that contribution of decreasing FWHM to mixing is not significant, in contrast the displacement of the Ti atoms across the interface is predominant contributing to the broadening of the mixed layers towards the substrate more than that in the opposite direction. The mixing amount is shown proportional to number of displaced atoms while the ion energy transfer to recoil may plays a major role in the degree of extending the mixed area.

*3.2. Ellipsometry results*

As mentioned above, after implantation the transition layers are broadened at the $TiO_2/SiO_2$ interface. On the basis of the ES measurements, the thickness of the layers including the mixed areas of $TiO_2/SiO_2$ structures obtained by RBS was confirmed. The MAIE (Multiple-angle-of-incidence Ellipsometry) method [17] was used for the calculations. There were considered four homogeneous layers: the $TiO_2$, the transition, $SiO_2$ layer and Si substrate. The figure below shows the model which describes the studied system before ion implantation. It was assumed that the interface between the layers is sharp and the layers are uniform. The optical constants of $TiO_2$ and $SiO_2$ were taken from the software of the VASE ellipsometer.

| 3 | *TiO₂ layer* | 30.8 nm |
|---|---|---|
| 2 | *TiO₂/ SiO₂ interface layer* | 3.2 nm |
| 1 | *SiO₂ layer* | 144.1 nm |
| 0 | *Si substrate* | 1 mm |

Fig.7. The model used in the Ellipsometry calculations.

The transition layer between $TiO_2$ and $SiO_2$ layers was described as a mixture of $TiO_2$ and $SiO_2$ compounds. The EMA (effective medium approximation) model was used in the calculations as Eq. (2). The detailed description of the application of the EMA method and its use in the calculations of optical constant material which is a mixture of two different materials can be found in [24, 25].

$$f_{TiO_2} \frac{\varepsilon_{TiO_2} - \varepsilon}{\varepsilon_{TiO_2} + 2\varepsilon} + (1 - f_{TiO_2}) \frac{\varepsilon_{SiO_2} - \varepsilon}{\varepsilon_{SiO_2} + 2\varepsilon} = 0 \qquad (2)$$

Based on the expression, the concentration of TiO$_2$ in the transition layer was calculated. The quantities in the formula mean ε dielectric function of the transition layer (of the mixture), $\varepsilon_{TiO_2}$ dielectric values of TiO$_2$, and $\varepsilon_{SiO_2}$ describe the SiO$_2$, $f_{TiO_2}$ is the concentration of TiO$_2$ in this mixture. The results of calculations made for the material before the ion implantation indicate that the transition layer of thickness 3.2 nm ± 0.3 nm contains 61.75 % TiO$_2$. These results are in good agreement with the data obtained by the RBS study. The thickness of the transition layers and the content of chemical compositions for the samples are shown in Table 2. It was assumed that the layers are heterogeneous mixed TiO$_2$ and SiO$_2$. Concentration of SiO$_2$ decreased with the growing ion energy, which could be associated with displacement of Ti atoms to the transition layer and a slow increase of thicknesses of the layers. The optical parameters of the first transition layers were also determined in these calculations.

Table 2. The concentration of chemical compounds and thickness of transition layers from SE obtained from the EMA model.

| Name of samples | RBS | Ellipstrometry spectroscopy | | |
|---|---|---|---|---|
| | Thickness of transition layers [nm] | Chemical composition [%] | | |
| | | SiO$_2$ | TiO$_2$ | mixture |
| Virgin | 3.0 | 3.2 ± 0.3 | 38.25 | 61.75 | 0.00 |
| Implanted E=100 keV | 10.3 | 10.0 ± 0.3 | 12.00 | 9.00 | 79.00 |
| Implanted E=150 keV | 11.9 | 10.7 ± 0.3 | 11.80 | 9.98 | 78.22 |
| Implanted E=200 keV | 12.7 | 11.2 ± 0.3 | 10.30 | 10.90 | 78.80 |
| Implanted E=250 keV | 12.7 | 12.1 ± 0.3 | 8.28 | 9.99 | 81.73 |

The results of Ellipsometric measurements of the spectra of Ψ(λ) and Δ(λ) for the TiO$_2$/SiO$_2$ structures are presented in Fig.8. For the non-implanted samples, the local extreme points are found in the range 309 nm to 317 nm and 545 nm to 561 nm for the spectra Ψ(λ) (Fig.8a). The other changes are observed in the same intervals in the Δ(λ) spectra (Fig.8b). The first extreme point observes in the Δ(λ) spectra around 309 nm is related to the phase change with changing the incidence angle of the light beam. Such waveforms measured at different incidence angles may suggest that the measurements are carried out near the main Principle point. This result suggests that the accuracy and sensitivity of the measurement with the help of SE method are at the maximum here. The band was found in the region from 560 nm to 650 nm, its yields vary with a change in the incidence angle and may be associated with the interference processes of light reflected from the examined structure. Similar investigations were carried out for the samples after ion implantation. The typical spectra collected from the samples implanted with the ions Xe energy of 250 keV results are presented in Fig 8.c, d. The value of local bands decreases with the increasing angle of incidence at which the spectra are collected. Changing the shape and width of the mentioned bands depending on the angle of incidence changes in the length of the path as the light beam overcomes during measurement can be explained.

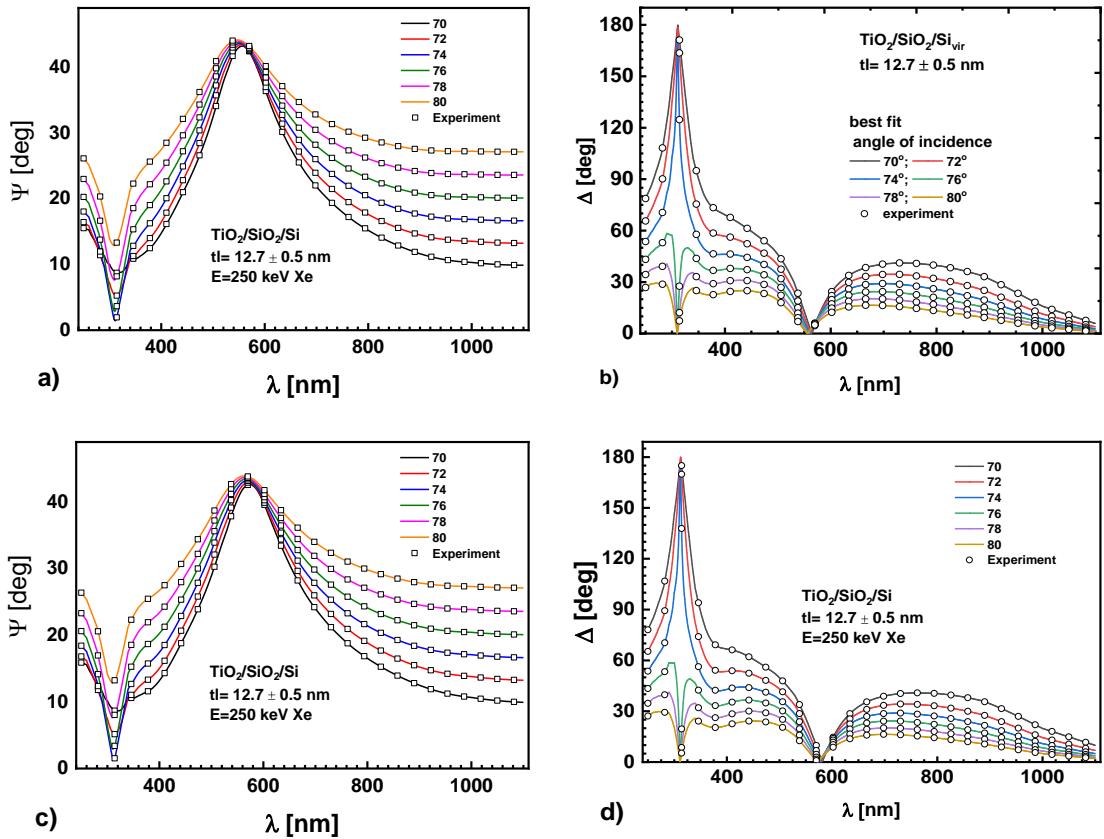

Fig.8. Spectra of Ψ(λ) and Δ(λ) ellipsometric angles measured for the samples before (a, b) and after (c, d) ion implantation. The solid lines present the best fit and the dot points are experiment results.

The spectra of the refractive index n and the extinction coefficient k as functions of the wavelength determined for the transition layers before and after implantation are presented in the figure below. It was observed that the parameters increase with the increasing wavelength up to 300nm (for n) and 260nm (for k), then decrease for in the remaining range from 300 to 1100nm as shown in Fig.9. For the virgin samples, the lowest values of n and k were obtained due to the mixing of atoms without ions and the disorders in this area. After implantation the n and k values increase in the range of Xe energy from 100 to 200 keV, then decrease for the 250 keV Xe ions. During irradiation, the transition layers changed the composition and the concentration of the elements as well as defects. This leads to the variation of the n and k values. As shown in the RBS results, the thickness of the transition layers increases with the increasing ion energy, indicating that the concentration of Xe in the layers grew with broadening of mixed area. As a result, the disorders produced by the incident ions increase and lead to the increasing absorption of $TiO_2$-$SiO_2$ and the possibility of light propagation in the mixed layers. For the highest energy of incident Xe ions, the projected range of the ions beyond the interface of $TiO_2$/$SiO_2$, the transition layer is less affected by collisions, thus the parameters n and k decrease to the values close to those for the case of 100 keV Xe implantation.

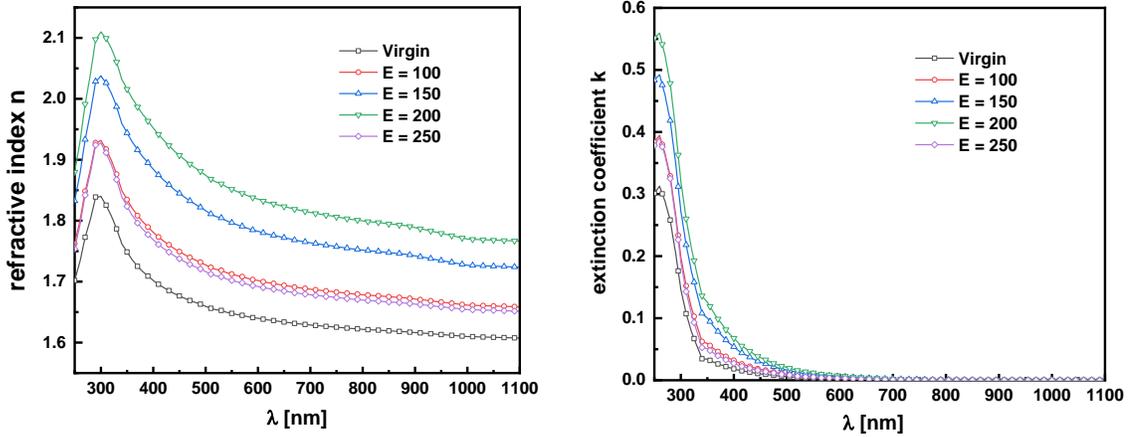

Fig.9. The typical spectra of the refractive index n (a) and extinction coefficient k (b) for the transition layers as functions of the wavelength.

## 4. Conclusions

In summary, Xe irradiation influences significantly on the $TiO_2/SiO_2$ bilayers for both structural and optical properties. The mixing induced by the ion beam at the $TiO_2$ and $SiO_2$ interfaces is pointed out by shifts of the borders of our measured RBS spectra. The amount of mixing which is qualified with FWHM difference of Ti peaks and relative thickness of transition layers, shown to be increase with increasing ion energy. These effects could be explained by increasing number and the path of the displacement atoms due to the increase the transferred energy of the ions in the transition layers. On the other hand, ellipsometric measurements of the spectra of $\Psi(\lambda)$ and $\Delta(\lambda)$ for the material before and after ion implantation show a correlation between in the yields of the bands and the angle of incident ion beam. This is associated with the interference processes of light reflected from the examined structure. The refractive index n and the extinction coefficient k have been found to increase after implantation up to 200 keV then to decrease at 250 keV Xe ion. Furthermore, the increasing of the thickness of transition layers between the $TiO_2$ and $SiO_2$ materials that has been observed in the ellipsometric measurements are in good agreement with the RBS results.

# References

[1] O. Aucillo, R. Kelly, 1984. *Ion Bombardment and Modification of Surfaces*, Elsevier, Amsterdam.

[2] J.W. Nastasi and M. Mayer, 2007, *Ion Beam Mixing*, in Radiation Effects in Solids, NATO Science Series, Springer, Dordrecht.

[3] L.S. Hung, M. Nastasi, J. Gyulai, J.W. Mayer, "Ion-induced amorphous and crystalline phase formation in Al/Ni, Al/Pd, and Al/Pt thin films", *Applied Physics Letters* 42: 672. https://doi.org/10.1063/1.94068.

[4] J.W. Mayer, B.Y. Tsaur, S.S. Lau, L-S. Hung, 1981, "Ion-beam-induced reactions in metal-semiconductor and metal-metal thin film structures", *Nuclear Instruments and Methods* 182-183: 1-13. https://doi.org/10.1016/0029-554X(81)90666-2.

[5] J.M. Poate, G. Foti, 1982, *Surface Modification and Alloying*, Plenum, New York. https://doi.org/10.1007/978-1-4613-3733-1.

[6] S. Matteson, M.A. Nicolet, 1983, "Ion Mixing", *Annual Review of Materials Science* 13: 339-362. https://doi.org/10.1146/annurev.ms.13.080183.002011.

[7] N. Bibic, S. Dhar, M. Milosavljevic, K. Removic, L. Rissanen, K.P. Lie, 2000, "Interface mixing in Ta/Si bilayers with Ar ions", *Nuclear Instruments and Methods in Physics Research B* 161–163: 1011–1015. https://doi.org/10.1016/S0168-583X(99)00838-1.

[8] S.K. Sinha, D.C. Kothari a, T. Som, V.N. Kulkarni, K.G.M. Nair, M. Natali, 2000, "Effects of Ne and Ar ion bombardment on Fe/$SiO_2$ bi-layers studied using RBS", *Nuclear Instruments and Methods in Physics Research B* 170: 120–124. https://doi.org/10.1016/S0168-583X(00)00068-9.

[9] M. Ni, M. K. H. Leung, D. Y. C. Leung and K. Sumathy, 2007, "A review and recent developments in photocatalytic water-splitting using $TiO_2$ for hydrogen production", *Renewable and Sustainable Energy Reviews* 11, 401–425. https://doi.org/10.1016/j.rser.2005.01.009.

[10] K. Nakata and A. Fujishima, '$TiO_2$ photocatalysis: Design and applications', 2012, *Journal of Photochemistry and Photobiology C* 13: 169–189. https://doi.org/10.1016/j.jphotochemrev.2012.06.001.

[11] G. Liu, L. Wang, H.G. Yang, H.-M. Cheng and G.Q. Lu, 2010, "Titania-based photocatalysts-crystal growth, doping and heterostructuring", *Journal of Materials Chemistry* 20: 831-843. https://doi.org/10.1039/B909930A.

[12] C. Anderson and A.J. Bard, 1997, "Improved Photocatalytic Activity and Characterization of Mixed $TiO_2/SiO_2$ and $TiO_2/Al_2O_3$ Materials" *The Journal of Physical Chemistry B* 101**:** 2611-2616. https://doi.org/10.1021/jp9626982.

[13] H. Tada, M. Akazawa, Y. Kubo, S. Ito, Enhancing Effect of $SiO_x$ Monolayer Coverage of $TiO_2$ on the Photoinduced Oxidation of Rhodamine 6G in Aqueous Media" *The Journal of Physical Chemistry B* 102 (1998) 6360-6366. https://doi.org/10.1021/jp980892d.

[14] S. Bharatkumar Patel, 2019, "A mechanical and modelling study of magnetron sputtered cerium-titanium oxide film coatings on Si (100)" *Ceramics International* 45**:** 6875-6844. https://doi.org/10.1016/j.ceramint.2018.12.183.

[15] G. Lassaletta, A. Fernández, J.P. Espinós, A.R. González-Elipe, 1995, "Spectroscopic characterization of quantum-sized $TiO_2$ supported on silica: influence of size and $TiO_2$-$SiO_2$ interface composition" *The Journal of Physical Chemistry* 99**,** 1484-1490. https://doi.org/10.1021/j100005a019.

[16] W.K. Chu, J.W. Mayer and M.A. Nicolet, 1978. *Backscattering Spectrometry*, Academic Press, New York.

[17] Hiroyuki Fujiwara, 2003. *Spectroscopic Ellipsometry: Principles and Applications*, John Wiley & Sons Ltd, England.

[18] M. Turek, S. Prucnal, A. Drozdziel and K. Pyszniak, 2011, "Plasma Ion Source with an Internal Evaporator" *Nuclear Instruments and Methods in Physics Research Section B* 269**:** 700–707. https://doi.org/10.1016/j.nimb.2011.01.133.

[19] M. Mayer, 1997. *SIMNRA User's Guide*, Report IPP 9/113, Max-Planck-Institut für Plasmaphysik, Garching, Germany.

[20] J.F. Ziegler, J.P. Biersack and M.D. Ziegler, 2008, *SRIM-the Stopping and Range of Ions in Matter, SRIM*, Ion Implantation Press, Chester.

[21] M. M. Ibrahim and N. M. Bashara, 1971, "Parameter-Correlation and Computational Considerations in Multiple-Angle Ellipsometry" *Journal of the Optical Society of America* 61**:** 1622-1629. https://doi.org/10.1364/JOSA.61.001622.

[22] Lawrence R.Doolittle, 1986, "A semiautomatic algorithm for Rutherford backscattering analysis", *Nuclear Instruments and Methods in Physics Research B* 15, 1–6, 1 (April): 227-231. https://doi.org/10.1016/0168-583X(86)90291-0.

[23] P. Sigmund, (1983), "Mechanism of ion beam induced mixing of layered solids" *Applied Physics A* 30: 43-46. https://doi.org/10.1007/BF00617712.

[24] D. E. Aspnes, J. B. Theeten, and F. Hottier, 1979, "Investigation of effective-medium models of microscopic surface roughness by spectroscopic ellipsometry" *Physical Review B* 20: 3292. https://doi.org/10.1103/PhysRevB.20.3292.

[25] D. E Aspnes, 1982, "Optical properties of thin films" *Thin Solid Films* 89**:** 249-262. https://doi.org/10.1016/0040-6090(82)90590-9.